# Thermal stress around a smooth cavity in a plate subjected to uniform heat flux

Zhaohang Lee[1], Yu Tang[1], Wennan Zou [*1]

(1. Institute of Engineering Mechanics / Institute for Advanced Study, Nanchang University, Nanchang 330031, China)

**ABSTRACT:** The two-dimensional thermoelastic problem of an adiabatic cavity in an infinite isotropic homogeneous medium subjected to uniform heat flux is studied, where the shape of the cavity is characterized by the Laurent polynomial. By virtue of a novel tactics, the obtained K-M potentials can be explicitly worked out to satisfy the boundary conditions precisely, and the possible translation of the cavity is also available. The new and explicit analytical solutions are compared with the those reported in literature and some serious problems are found and corrected. Finally, some discussions on the thermal stress concentration around the tips of three typical cavities are provided.

## 0 Introduction

Plates under the environment of changing temperature are extensively used in designing new steam and gas turbines, high speed flight vehicles, jet and rocket engines, nuclear reactor, and various machine structures, and also in the fields of nuclear and chemical engineering. Studies on stress concentration due to the presence of a cavity in an infinite plate have become a topic of considerable research (Savin, 1961; Kattis, 1991; Zou and He, 2018), as well as the corresponding studies cited therein. Under the action of uniform heat flux, if there is no constraint in the plate and the material deforms freely, there will be no stress. However, the continuous deformation of the material must be interrupted due to the appearance of cavity, which would be necessary at the beginning of design or the openings required by the design. This is called the thermal stress around the cavity. When the stress level is higher than the ultimate strength of the material, it will lead to structural failure. Determining the thermal stress around the cavity can effectively predict and evaluate the properties of materials.

The K-M potentials established by Muskhelishvili (1953) provide a powerful tool for the plane problem of isotropic elasticity. Two analytic complex functions as the K-M potentials are introduced to expressed the displacement and the stress in order to naturally satisfy the constitutive relations and the equibrium equations, and the original problem is then transformed into the boundary value problem of analytic functions. Due to the convenience and universality of the K-M potential method, it is also applied to the thermalelastic problem involved in this paper.

The problem to determine the thermal stress around a non-elliptical cavity with the adiabatic boundary in an isotropic plate under uniform heat flux has been considered for a long time. Florence and Goodier (1960) studied the thermal stress around an ovaloid hole and analysed two specsified cases of the elliptical hole and the slot. Deresiewicz (1961) extended the study to holes with general shapes described by the Laurent polynomials, and derived the explicit solutions through adding the counter terms with positive powers. Years later, Yoshikawa and Hasebe (1999) studied the thermal elasticity of arbitrary shape holes in infinite media when a point heat source is at any position in the plane. Bhullar and Wegner (2009) analyzed the thermal stress of the hyperelliptical hole under the uniform heat flow by using complex variable method under isothermal conditions. They found that the stress concentration at the tip of the hole is very serious. Jafari et al. (2016a) analyzed the thermal stress distribution of hypocycloidal holes. Jafari *et al.* (2016b) studied the thermal stress of triangular holes with different shape parameters and heat flux directions. Chao *et al.* (2018) presented a series solution of the thermoelastic problem of triangular holes with coating. Yu *et al.* (2019) studied the elastic problems of a thermoelectric material containing an arbitrarily-shaped hole under a uniform remote electric current and a uniform energy flux. Tseng *et al.* (2020) studied the case of a square hole with coating. It is found that (1) the non-elliptical shapes of cavities the researchers studied, with few exceptions, are those characterized by the Laurent polynomials, (2) there is no significant progress in the analytic solution since Deresiewicz (1961), most of the research works are about applications and extensions.

---
[1] Zhaohang Lee, email address: 351113619002@email.ncu.edu.cn; Corresponding author: Wennan Zou, email address: zouwn@ncu.edu.cn.

In this paper, based on the tactics proposed in our previous work (Zou and He, 2018), a new solution for the thermal stress problem of uniform heat flux applied in an isotropic plate with a non-elliptical cavity under the adiabatic boundary condition is obtained. This solution is more operable and effective than that given by Deresiewicz (1961). The rest of this paper is arranged as follows. In Section 2, the thermal stress problem is briefly formulated, where the temperature distribution under remote uniform heat flux is derived in a compact process, and the thermal stress in the context of the K–M potential theory is divided in two parts: one is to balance the thermal dislocation and consider the relative rigid-body translation of the cavity to the matrix, another is about the perturbance reducing to zero at infinity and guranteeing the satisfaction of traction-free condition on the boundary of the cavity. The basic potentials described the thermal dislocation under an isothermal state, but without regard to the boundary loading, are first proposed by Florence and Goodier (1960). In second 3, the general explicit potentials, never reported before, are presented when the shape of the cavity is characterized by a Laurent polynomial while postponing the detailed derivations to Appendix A, the effectiveness of the new solution is discussed by comparing it with the previous results. Aanlyses of stress distribution are expanded in Section 4: three typical shapes, triangle, square and pentagram star, are considered; hydrostatic pressure, maximal shear stress, and toroidal normal stress along the contour of the cavity are illustrated; the relation between the stress concentration and curvature, the effect of heat flux direction, and the decay of stress around the tips are discussed. Some concluding remarks are drawn in Section 5.

## 2 Basic equations

### 2.1 Description of the problem

Consider an infinite body $\Omega$ in two-dimensional space consisting of a homogeneous and isotropic medium whose thermal conduction behaviour is governed by Fourier's law, and elastic behaviour by Hooke's law. We are concerned with the perturbance effect due to a free cavity with a traction-free, thermally insulated boundary while the matrix is subjected to uniform heat flux at infinity, as shown in Fig. 1. By the Riemann mapping theorem in complex analysis, there is a unique function in the form of Laurent series (see, e.g., Zou *et al.* 2010)

$$z = z(w) = h + R\phi(w) = h + R\left(w + \sum_{k=1}^{\infty} b_k w^{-k}\right), |w| \geq 1, \tag{1}$$

mapping the exterior of the cavity onto the exterior of the unit circle with the origin as its center. In the above expression, $h$ is a point inside the cavity, $R$ is a positive real parameter indicating the size of the cavity, and $b_k$ are the complex variable parameters representing the shape. It can be seen that the point $t$ on the boundary of any simply connected shape can be accurately described by $t = z(\eta), |\eta| = 1$. In general, we only need to take a limited number of terms $N$ to meet the accuracy requirement. To continue to improve the accuracy, we only need to increase the number of terms. So, one might as well use

$$t = h + R\phi(\eta) = h + R\left(\eta + \sum_{k=1}^{N} b_k \eta^{-k}\right), |\eta| = 1 \tag{2}$$

to describe the cavity and for the problem of a sole cavity, $h = 0$ is usually taken.

The uniform heat flux at infinity is described by

$$\boldsymbol{q}^\infty = q\boldsymbol{n}_\beta, \tag{3}$$

where $\boldsymbol{n}_\beta$ is the unit vector indicating the direction of the heat flux, as shown in Fig. 1, and can be denoted by a complex variable $n_\beta = e^{\iota\beta}$, with $\beta$ being the included angle between the heat flux direction and the *x*-axis. It is assumed that the deformation under the action of thermal stress is always infinitesimal in the linear elastic range. In this paper, the Cartesian coordinate system is adopted, with $(x_1, x_2)$ indicating an arbitrary point and $z = x_1 + \iota x_2$, $\iota = \sqrt{-1}$ the corresponding complex variable.

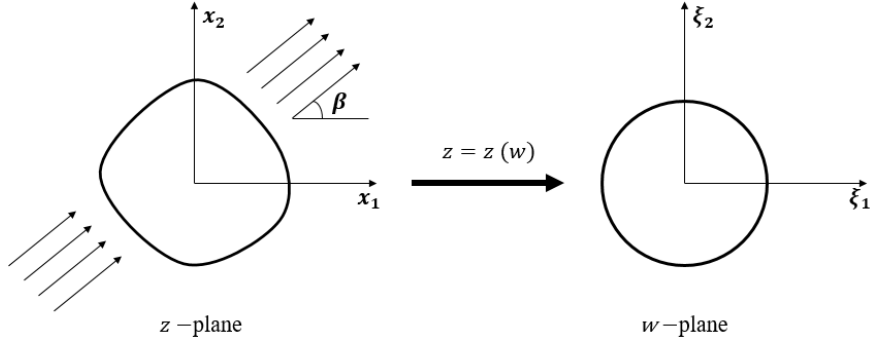

**Fig. 1.** Conformal mapping of the cavity.

## 2.1 Temperature

The stationary temperature field of linear and an isotropic thermal material is a harmonic function satisfying the Laplacian equation $\nabla^2 T = 0$. Denote the negative temperature gradient by

$$\boldsymbol{e} = -\nabla T, \tag{4}$$

we can write the heat flux as

$$\boldsymbol{q} = -k\nabla T = k\boldsymbol{e}, \tag{5}$$

where $k$ is the thermal coefficient of the matrix. Besides the remote condition (3), the adiabatic condition across the boundary of the cavity is

$$q_n = \boldsymbol{q} \cdot \boldsymbol{n} = 0, \tag{6}$$

where $\boldsymbol{n}$ is the outer normal direction of the boundary.

Now introduce the complex representations, say the direction vector $n = n_1 + \iota n_2$. The complex temperature function $H(z) = T + \iota T^c$ is constructed by using the temperature field $T$ and its conjugate harmonic function $T^c$, which is an analytic function outside the cavity and satisfies the Cauchy-Riemann relations:

$$\partial_1 T = \partial_2 T^c, \quad \partial_2 T = -\partial_1 T^c, \tag{7}$$

where $\partial_i$ represents the partial derivative of the coordinate $x_i$. Substitution of (7) into (4) and (5) yields

$$q = ke = -k(\partial_1 T + \iota \partial_2 T) = -k\overline{H'(z)}, \tag{8}$$

where "$\overline{(\cdot)}$" denotes the conjugation of a complex variable $(\cdot)$. Thus, from

$$q_n = -k\mathbf{n} \cdot \nabla T = -k(n_1 \partial_1 T + n_2 \partial_2 T) = -k\mathrm{Re}[n(\partial_1 T - \iota \partial_2 T)] = -k\mathrm{Re}[nH'(z)], \tag{9}$$

$$q^\infty = -k\overline{H'(z)}\big|_{z\to\infty}. \tag{10}$$

where Re[·] indicates the real part of a complex variable, the constraint conditions (3) and (6) can be expressed by

$$nH'(t) + \overline{nH'(t)} = 0, \text{ on } t = z(\eta), |\eta| = 1; \tag{11}$$

$$H'(z) = -\frac{q}{k}e^{-\iota\beta}, \text{ on } z \to \infty. \tag{12}$$

The analytic function $H(z)$ satisfying (11) and (12) gives the solution of temperature problem.

By virtue of the expression

$$n = -\iota \frac{dt}{ds} \tag{13}$$

on the boundary of the cavity, where *s* is the arc length coordinate along the boundary, substituting it into (11) to do an integral along the boundary gives

$$H(t) - \overline{H(t)} = \text{const.}, \text{ on } t = z(\eta), |\eta| = 1. \tag{14}$$

Following Hasebe and Tamai (1986), the complex temperature can be broken down into two parts
$$H(z) = H_1(z) + H_2(z) \tag{15}$$
where the basic part $H_2(z)$ is determined by the reference temperature $T_0$ of the material without thermal stress and the remote heat flux $q$ as
$$H_2(z) = T_0 - \frac{q}{k} z e^{-\iota\beta}, \tag{16}$$
while $H_1(z)$ is the complementary and holomorphic part due to the cavity, satisfying
$$H_1(t) - \overline{H_1(t)} = \frac{q}{k}\left(t e^{-\iota\beta} - \bar{t} e^{\iota\beta}\right), \text{on } t = z(\eta), |\eta| = 1; \tag{17}$$
$$H_1(z) = H_1'(z) = 0, \text{on } z \to \infty. \tag{18}$$
Considering the holomorphic property of $H_1(z)$ at infinity in the image plane and the mapping function (2), we obtain the solution
$$H_1(z(w)) = H_1(w) = \sum_{k=1}^{\infty} a_k w^{-k} = \frac{qR}{k}\left(e^{-\iota\beta} \sum_{k=1}^{N} b_k w^{-k} - w^{-1} e^{\iota\beta}\right), |w| \geq 1 \tag{19}$$
in terms of the complex variable $w$ in the image plane.

Finally, substitution of (16) and (19) into (15) yields the complex temperature field
$$H(z(w)) = T_0 - \frac{qR}{k}\left(w e^{-\iota\beta} + w^{-1} e^{\iota\beta}\right). \tag{20}$$
This result was first reported by Florence and Goodier (1960) for the problem of an insulated ovaloid cavity, and directly used to the general shapes (1) by Deresiewicz (1961) without any explanation. But Jafari *et al.* (2016a, 2016b) presented wrong expression of the complex temperature.

**2.2 Thermal stress**

According to the theory of planar elasticity established by Muskhelishvili (1963), two analytic functions $\varphi(z)$ and $\psi(z)$, called the Kolosov-Muskhelishvili (K-M) potentials, can be introduced to indicate the stress and the displacement of point 2 relative to point 1 as
$$\begin{cases} \sigma_{11} + \sigma_{22} = 2[\varphi'(z) + \overline{\varphi'(z)}], \\ \sigma_{22} - \sigma_{11} + 2\iota\tau_{12} = 2[\bar{z}\varphi''(z) + \psi'(z)], \end{cases} \tag{21}$$
$$u_1 + \iota u_2 = \frac{1}{2\mu}\left[\kappa\varphi(z) - z\overline{\varphi'(z)} - \overline{\psi(z)}\right] + \alpha' \int_1^2 \text{Re}[H(y) - T_0] dy, \tag{22}$$
where the effect of thermal expansion is taken into account, $\mu$ is the shear modulus of the material, $\kappa$ and $\alpha'$ are parameters associated with Poisson's ratio $\nu$ and linear expansion coefficient $\alpha$, respectively, say
$$\kappa = \begin{cases} \frac{3-\nu}{1+\nu}, \text{plane stress,} \\ 3-4\nu, \text{plane strain;} \end{cases} \alpha' = \begin{cases} \alpha, \text{plane stress,} \\ (1+\nu)\alpha, \text{plane strain.} \end{cases} \tag{23}$$
The stress in the above formulae naturally satisfies the compatible relations and the equilibrium equation without body force. The traction $f^n = f_1^n + \iota f_2^n$ on the surface with normal $n$ and arc length coordinate $s$ is described by
$$f^n = -\iota \frac{d}{ds}\left[\varphi(t) + t\overline{\varphi'(t)} + \overline{\psi(t)}\right]. \tag{24}$$

The thermal dislocation (Florence and Goodier, 1960) coming from the temperature (20) needs basic K-M potentials in the form
$$\varphi_0(z(\eta)) = A\ln\eta, \quad \psi_0(z(\eta)) = 2\mu\overline{U_0} + B\ln\eta, \tag{25}$$
where $A$ and $B$ are two undetermined parameters with dimension of linear force, $U_o$ indicates the induced rigid-body translation of the cavity relative to the matrix (Zou and He, 2018). The remaining perturbation potentials admit the representation
$$\varphi_p(z(w)) = R\sum_{k=1}^{\infty} \alpha_k w^{-k}, \quad \psi_p(z(w)) = R\sum_{k=1}^{\infty} \beta_k w^{-k}, \quad |w| \geq 1. \tag{26}$$
using the complex variable $w$ in the image plane, where $\alpha_k$ and $\beta_k$ are two sets of coefficients with dimension of traction. Sum of (25) and (26) gives the total K-M potentials as

$$\varphi(z) = \varphi_0(z) + \varphi_p(z), \qquad \psi(z) = \psi_0(z) + \psi_p(z). \tag{27}$$

Taking account of the property $H(t) = \overline{H(t)}$ on the boundary and the origin inside the cavity, the total dislocation around the cavity anticlockwise can be calculated from (22), and should vanish such that

$$\Delta_U = [u_1 + \iota u_2]_{\eta=e^{0\iota}}^{e^{2\pi\iota}} = \frac{1}{2\mu}\left[\kappa\varphi(t) - t\overline{\varphi'(t)} - \overline{\psi(t)}\right]_{\eta=e^{0\iota}}^{e^{2\pi\iota}} + \alpha' \oint H(t)dt = 0. \tag{28}$$

Substitution of (25)-(27) yields

$$\kappa A + \bar{B} = \frac{2\mu\alpha' q R^2}{k}\left(e^{\iota\beta} - b_1 e^{-\iota\beta}\right). \tag{29}$$

In addition, the resultant force on the boundary of the cavity should keep in balance, from (24), that means

$$\Delta_F = \frac{1}{2\mu}\left[A\ln\eta + \frac{\bar{A}z(\eta)}{\overline{z'(\eta)}}\eta + \overline{B\ln\eta} + 2\mu\overline{U_0}\right]_{\eta=e^{0\iota}}^{e^{2\pi\iota}} = \frac{\pi\iota}{\mu}(A - \bar{B}) = 0, \tag{30}$$

Combination of (29) and (30) yields the solution

$$A = \bar{B} = \frac{2\mu\alpha' q R^2}{k(\kappa+1)}\left(e^{\iota\beta} - b_1 e^{-\iota\beta}\right). \tag{31}$$

It can be found that $A$ is a parameter associated with the material properties, the heat flux and the shape characteristic of the cavity. It is remarkable that only two effective shape parameters, $R$ and $b_1$, enter the formula of $A$, and particularly the thermal dislocation disappears when the cavity becomes a slot parallel to the direction of the heat flux, namely $b_1 = e^{2\iota\beta}$. It is found that the formula of $A$ given by Jafari *et al.* (2016a, 2016b) is also wrong. Since the potentials have dimension of linear force, no loss of generality, in this paper, we define the characteristic value of stress as

$$\sigma_0 = \frac{2\mu\alpha' q R}{k(\kappa+1)}, \tag{32}$$

and so

$$A = R\sigma = R\sigma_0\left(e^{\iota\beta} - b_1 e^{-\iota\beta}\right). \tag{33}$$

From (24), the boundary condition for traction-free can be expressed by

$$\varphi(t) + t\overline{\varphi'(t)} + \overline{\psi(t)} = 0. \tag{34}$$

Substituting of (25), (27) and (31) yields the constraint condition of the perturbance potentials on the boundary

$$\varphi_p(\eta) + \frac{z(\eta)}{z'(\eta)}\overline{\varphi'_p(\eta)} + \overline{\psi_p(\eta)} = -2\mu U_0 - \frac{\bar{A}z(\eta)}{\overline{z'(\eta)}}\eta, |\eta| = 1, \tag{35}$$

which will be used to solve $\varphi_p(\eta)$ and $\psi_p(\eta)$.

## 3 General explicit solution and its effectiveness

Following the method proposed by Zou and He (2018), we obtain the following results: (1) the perturbance potentials have finite expression, namely $\varphi_p(\eta)$ has maximal negative power $N - 1$ while $\psi_p(\eta)$ multiplied by $\phi'(\eta)$ has maximal negative power $N + 1$; (2) the total K-M potentials for the problem of an insulated cavity characterized by (2) can be worked out in the form

$$\varphi(w) = A\ln w + R\sum_{k=1}^{N-1}\alpha_k w^{-k}, \qquad \psi(w) = \bar{A}\ln w + 2\mu\bar{U}_0 + \frac{R}{\phi'(\eta)}\sum_{k=1}^{N+1}\beta_k w^{-k}, \tag{36}$$

where $A$ is given by (33), $\alpha_k$ are solved from the linear equations (A.9), the rigid-body translation $U_0$ is gotten from (A.10) as

$$U_0 = \frac{R}{2\mu}\left(\sum_{l=1}^{N-1} l g_{l+1} \bar{\alpha}_l - \bar{\sigma} g_1\right)., \tag{37}$$

where the shape parameters $g_k$ are directly calculated by (A.3), $\beta_k$ are calculated by

$$\sum_{k=1}^{N+1}\beta_k w^{-k} = \sum_{l=k}^{N} l(b_l\bar{\alpha}_{l-k+1} + \alpha_l\bar{b}_{l-k+1})w^{-k} + \sum_{k=1}^{N-1} k\alpha_k w^{-k-2} - \sigma w^{-2} + \frac{2\mu\bar{U}_0}{R}[1 - \phi'(w)]. \tag{38}$$

The detailed derivation is presented in Appendix A.

The problem of thermal stress involved in this paper is classical and important. For the cases of a cavity with the shape characterized by the Laurent polynomial (2), Florence and Goodier (1960) pioneered the study of thermal stress distribution caused by the thermal dislocation, and dealt with the first non-elliptical cavity of ovaloid form. Deresiewicz (1961) extended to cavities whose boundary can be described by the Laurent polynomials, and presented a general solution for arbitrary shapes containing terms with positive powers. Since then, there has been no significant progress. For instance, the solutions given by Jafari *et al.* (2016a, b) completely followed the method of Deresiewicz (1961). The solutions (36)-(38) present a new form that contain no terms with positive powers. In the following, for comparison, we calculate the resultant force on the boundary of the cavity according to the solutions of the previous studies and our formulae, namely, testing whether the residual of TBC equal to zero everywhere:

$$\text{TBC} = \varphi(\eta) + \frac{z(\eta)}{\overline{z'(\eta)}} \overline{\varphi'(\eta)} + \overline{\psi(\eta)} = 0, \eta = e^{\iota\theta}, 0 \leq \theta < 2\pi. \tag{39}$$

Two examples are listed for comparison in the following.

(1). For a cavity characterized by $t(\eta) = R\left(\eta + \frac{m}{\eta} + \frac{n}{\eta^3}\right), |\eta| = 1$, the reported results (Florence and Goodier, 1960) are

$$\begin{cases} \varphi(w) = A\ln w - \bar{A}nw^{-2}; \\ \psi(w) = \bar{A}\ln w + Anw^2 - (Aw^2 + 2\bar{A}n)\dfrac{1 + mw^2 + nw^4}{w^4 - mw^2 - 3n}, \end{cases} \tag{40}$$

where a typo is corrected by changing the sign before $\bar{A}$ in the perturbance potential $\psi_p(w)$. The solutions of ours can be expressed with $A$, $b_1$ and $b_3$ as

$$\begin{cases} \varphi(w) = A\ln w - \bar{A}b_3 w^{-2}; \\ \psi(w) = \bar{A}\ln w + 2\mu\bar{U}_0 + \dfrac{R\beta_2 w^{-2} + R\beta_4 w^{-4}}{\phi'(w)}. \end{cases} \tag{41}$$

With

$$2\mu U_0 = -\bar{A}(b_1 + \bar{b}_1 b_3) - 2Ab_3\bar{b}_3$$
$$R\beta_2 = -A(b_1\bar{b}_1 + b_1^2 \bar{b}_3 + 3b_3\bar{b}_3 + 1) - 2\bar{A}b_3(\bar{b}_1 + b_1\bar{b}_3), \tag{42.1}$$
$$R\beta_4 = -3Ab_3(\bar{b}_1 + b_1\bar{b}_3) - 2\bar{A}b_3(1 + 3b_3\bar{b}_3). \tag{42.2}$$

Let $b_1 = m$, $b_3 = n$, we can check that the formulae (40) and (41) are exactly the same, but the constant part in $\psi(w)$ is now given a clear physical meaning. For the special case $R = 1, \beta = 0, m = 0.2, n = -0.05$, the residuals scaled of TBC by $R\sigma_0$ all equal to zero well, as shown in Fig. 2(a).

(2). Deresiewicz (1961) reported the potentials of triangular cavities characterized by $t(\eta) = R(\eta + c_2\eta^{-2} + c_5\eta^{-5}), |\eta| = 1$, as

$$\begin{cases} \varphi(w) = A\ln w - (\bar{A}\alpha_1 w^3 + Ac_5\alpha_1 w + \bar{A}c_5)w^{-4}, \\ \psi(\eta) = \bar{A}\ln w + Ac_5 w^4 + \bar{A}c_5\alpha_1 w^3 + A\alpha_1 w + (Aw^4 + \bar{A}\alpha_1 w^3 + 3Ac_5\alpha_1 w + 4\bar{A}c_5)\dfrac{1 + c_2 w^3 + c_5 w^6}{5c_5 + 2c_2 w^3 - w^6}, \end{cases} \tag{43}$$

with $\alpha_1 = \dfrac{c_2(1+2c_5)}{1-3c_5^2}$. According to our formulae, the potentials can be obtained to be

$$\begin{cases} \varphi(w) = A\ln w - R(\bar{\sigma}\alpha w^{-1} + \sigma b_5 \bar{\alpha} w^{-3} + \bar{\sigma}b_5 w^{-4}), \\ \psi(w) = \bar{A}\ln w + 2\mu\bar{U}_0 + R\dfrac{\beta_2 w^{-2} + \beta_3 w^{-3} + \beta_5 w^{-5} + \beta_6 w^{-6}}{\phi'(w)}, \end{cases} \tag{44}$$

with $\alpha = \dfrac{b_2 + 2\bar{b}_2 b_5}{1 - 3b_5\bar{b}_5}$, and

$$\frac{2\mu}{R}U_0 = -\sigma\bar{\alpha}(b_2 + 2\bar{b}_2 b_5) - 4\sigma b_5\bar{b}_5; \beta_2 = -\sigma(1 + 5b_5\bar{b}_5 + 3\bar{\alpha}b_5\bar{b}_2 + 2\bar{\alpha}b_2), \tag{45.1}$$

$$\beta_3 = -\bar{\sigma}\alpha(1 + 5b_5\bar{b}_5) - 4\bar{\sigma}b_5\bar{b}_2 + \frac{4\mu\bar{U}_0 b_2}{R}, \beta_5 = -8\sigma\bar{\alpha}b_5, \beta_6 = -4\bar{\sigma}b_5 + \frac{10\mu\bar{U}_0 b_5}{R}. \tag{45.2}$$

It can be verified that our results are the same as (42) by taking $b_2, b_5$ to be real and letting $A = R\sigma, b_2 = c_2, b_5 = c_5$. For the special case, $R = 1, \beta = 0, c_2 = \frac{1}{3}, c_5 = \frac{1}{45}$, the residuals of TBC scaled by $R\sigma_0$ are tested and shown in Fig. 2(b).

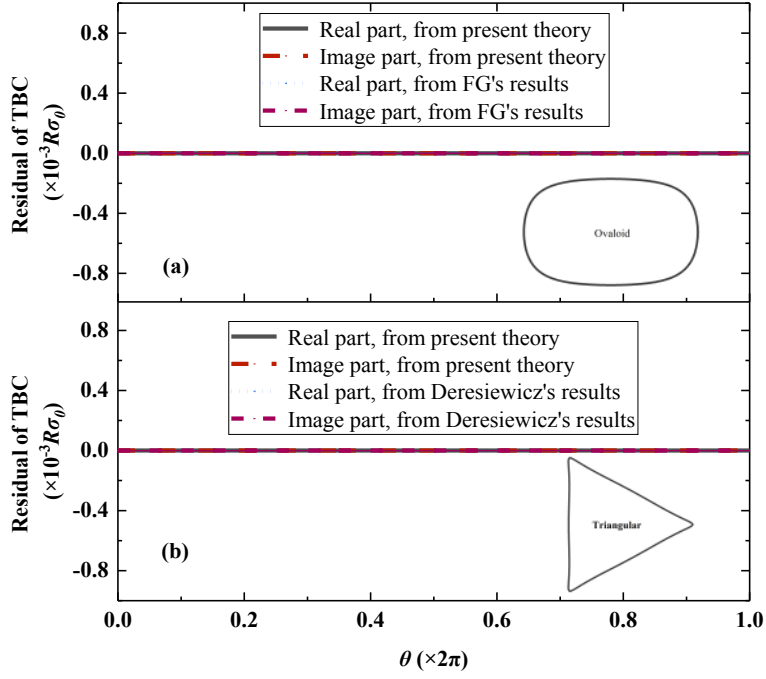

**Fig. 2** Resultant force residual comparison between FG's results and Deresiewicz 's results with ours, scaled by $R\sigma_0$ on the boundary of the cavities.

The above comparison make clear that Florence and Goodier's (FG's), Deresiewicz's solutions and ours are all correct near the boundary, but the constant term in the second potential is out of their initial construction and we point out the physical meaning of this term as the possible rigid-body translation between the matrix and the cavity. In addition, making use of terms with positive powers by Deresiewicz must result in the operation problem of big numbers when the shapes need terms of high degrees to characterize. As an example, the cavity of regular dodecagon described by

$$t(\eta) = R\left(\eta + \frac{1}{66}\eta^{-11} + \frac{55}{18216}\eta^{-23}\right), |\eta| = 1 \tag{46}$$

with $R = 1$ is considered. From (21), the hydrostatic pressure

$$\sigma_h = \frac{1}{2}(\sigma_{11} + \sigma_{22}) = 2\text{Re}[\varphi'(z)] \tag{47}$$

and the maximal shear stress (MSS)

$$\tau_{max} = \sqrt{\left(\frac{\sigma_{11} - \sigma_{22}}{2}\right)^2 + \sigma_{12}^2} = |\bar{z}\varphi''(z) + \psi'(z)| \tag{48}$$

can be formulated and the results from the present theory and Deresiewicz's are illustrated in Fig. 3. It is easy to see that the maximal shear stress field of Deresiewicz's solution is heavily contaminated for points away from the boundary but the hydrostatic pressure fields of the present theory and Deresiewicz's are the same. This is because the calculation of maximal shear stress needs the second potential which involves some terms of positive powers in Deresiewicz's theory.

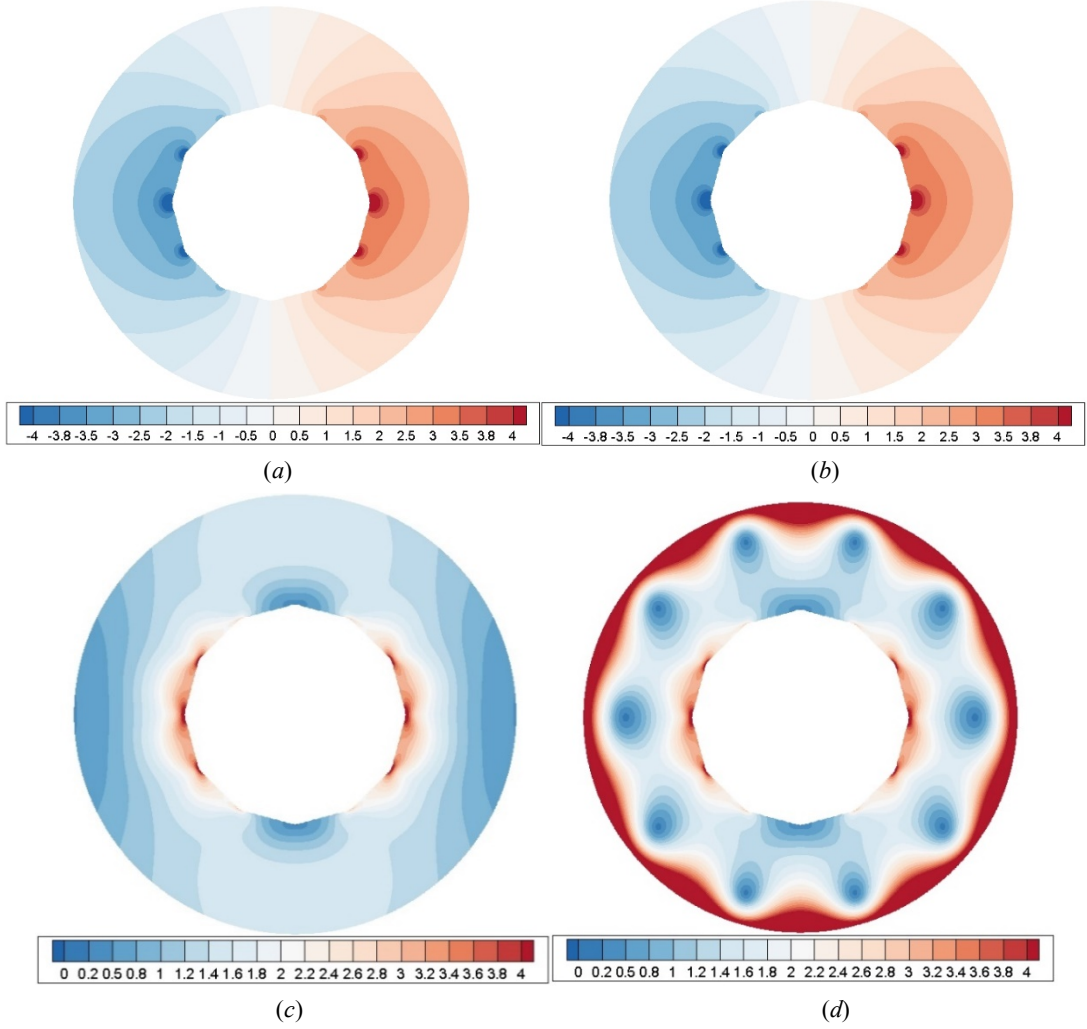

**Fig. 3.** Big number problem of a cavity of regular dodecagon in Deresiewicz's solution, all stresses scaled by $\sigma_0$ in (32), the scale of the calculation domain is about $2R$: (*a*, *b*) hydrostatic pressure of present theory (left) and Deresiewicz's (right); (*c*, *d*) maximal shear stress of present theory (left) and Deresiewicz's (right).

## 4 Results and analyses of stress distribution

Under the action of remote uniform heat flux, stress concentration will appear near the cavity, and the stress concentration at the tip is more obvious and serious, which is the most prone position for material failure. This property has an important impact on industrial design and material performance. Therefore, we will focus on the stress distribution at the tip in this section. Since the distribution of thermal stress at the tip of a cavity varies with the shape and the direction of heat flux, especially when the cavity has multiple tips, the stress distribution at each tip is also different. Based on the above understanding, we calculate the toroidal normal stress around the boundary of the cavity, and the hydrostatic pressure and maximum shear stress near the tip of different cavities and/or tips, in order to discuss the influence of heat flux direction on the stress around each tip. For convenience, the size parameter $R$ of different shapes is chosen to guarantee the cavities in comparison have the same area. All lengths are scaled by $R$, and all stresses by $\sigma_0$. We choose triangle, square and pentagram star as three representative shapes, the expressions are listed as follows (more items can be added according to the accuracy requirements):

$$t(\eta) = R\left(\eta + \frac{1}{3}\eta^{-2} + \frac{1}{45}\eta^{-5} + \frac{1}{162}\eta^{-8}\right), R = 0.64087, \tag{49.1}$$

$$t(\eta) = R\left(\eta + \frac{1}{6}\eta^{-3} + \frac{1}{56}\eta^{-7} + \frac{1}{176}\eta^{-11}\right), R = 0.59011, \tag{49.2}$$

$$t(\eta) = R\left(\eta + \frac{3\iota}{10}\eta^{-4} + \frac{13}{225}\eta^{-9} - \frac{4\iota}{125}\eta^{-14} - \frac{214}{11875}\eta^{-19} + \frac{1231\iota}{93750}\eta^{-24}\right.$$
$$+ \frac{20974}{2265625}\eta^{-29} - \frac{2908\iota}{390625}\eta^{-34} - \frac{441199}{76171875}\eta^{-39} + \frac{95763\iota}{19531250}\eta^{-44} \quad (49.3)$$
$$\left.+ \frac{6890609}{1708984375}\eta^{-49}\right), R = 0.74224,$$

with $|\eta| = 1$. These mapping functions for cavity shapes with tip indices are drawn in Fig. 5 (*a*)-(*c*), respectively. Two heat flux directions $\beta = 0$ and $\beta = \frac{\pi}{2}$ are considered for the distribution of maximal shear stress.

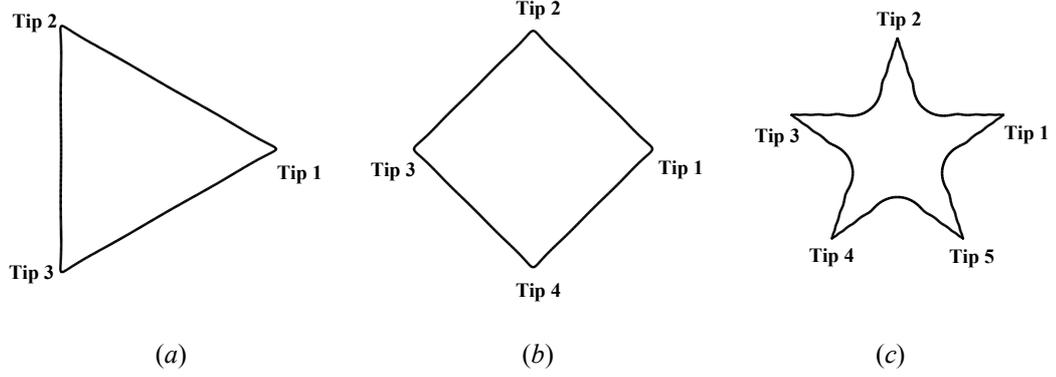

(*a*)          (*b*)          (*c*)

**Fig. 4.** Three cavities approximated by Laurent polynomials: (*a*) triangle, (*b*) square, (*c*) pentagram.

### 4.1 Maximal shear stress (MSS) around the cavities

As shear failure being the main failure mode of metal materials, the distribution of maximum shear stress (MSS) near the cavity is an important topic in strength analysis. Fig. 5 (*a*)-(*c*) and (*d*)-(*f*) show the fields of MSS near different cavities when the heat flux has directions $\beta = 0$ and $\beta = \pi/2$, respectively. It can be found that the large stress is mainly distributed around the cavities, and there are different degrees of stress concentration at the tip, and the MSS at each tip is not independently distributed, but permeates each other. Regardless of $\beta = 0$ or $\beta = \pi/2$, the most obvious position of stress concentration (the position that is most unfavorable to industrial design and material performance) is located near the tip(s) whose symmetry axis direction has the smallest angle with the heat flux direction, and the distribution of MSS is found be symmetrical. In order to compare the severity of stress concentration in different shapes of cavities, the tips with the most obvious stress concentration, say Tip1 when $\beta = 0$ and Tip2 when $\beta = \pi/2$ are selected to investigate with a magnifying glass, as shown in Fig. 6. Due to the smoothness of conformal mapping, the tip of pentagonal star is composed of a platform, namely has two points with the maximal curvature. It can be clearly seen that the severity of stress concentration is pentagonal > triangle > square; all the maximal MSSs appear at the points with maximal curvature except the case of Tip2 of triangle cavity when $\beta = \pi/2$. The case of the point with the maximum MSS deviating from the maximum curvature point will be further explained later.

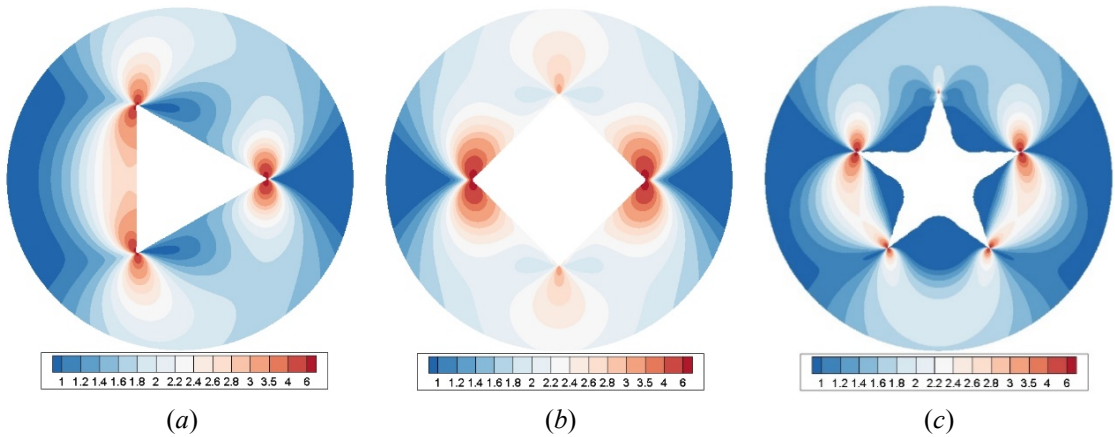

(*a*)          (*b*)          (*c*)

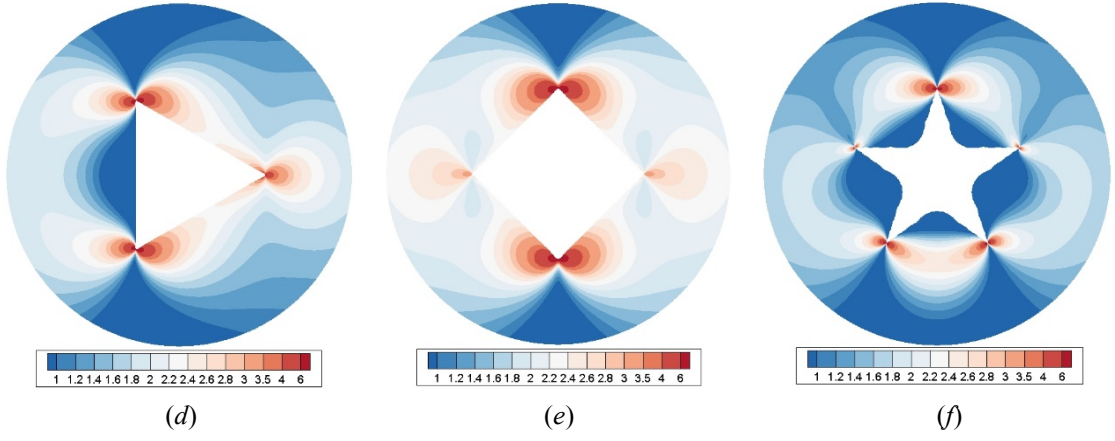

**Fig. 5.** Maximal shear stress scaled by $\sigma_0$ in (32) for various cavities under uniform heat flux: (*a-c*). the heat flux direction is $\beta = 0$; (*d-f*). the heat flux direction is $\beta = \pi/2$. The computational domain is a circular domain whose radius is two times $R$.

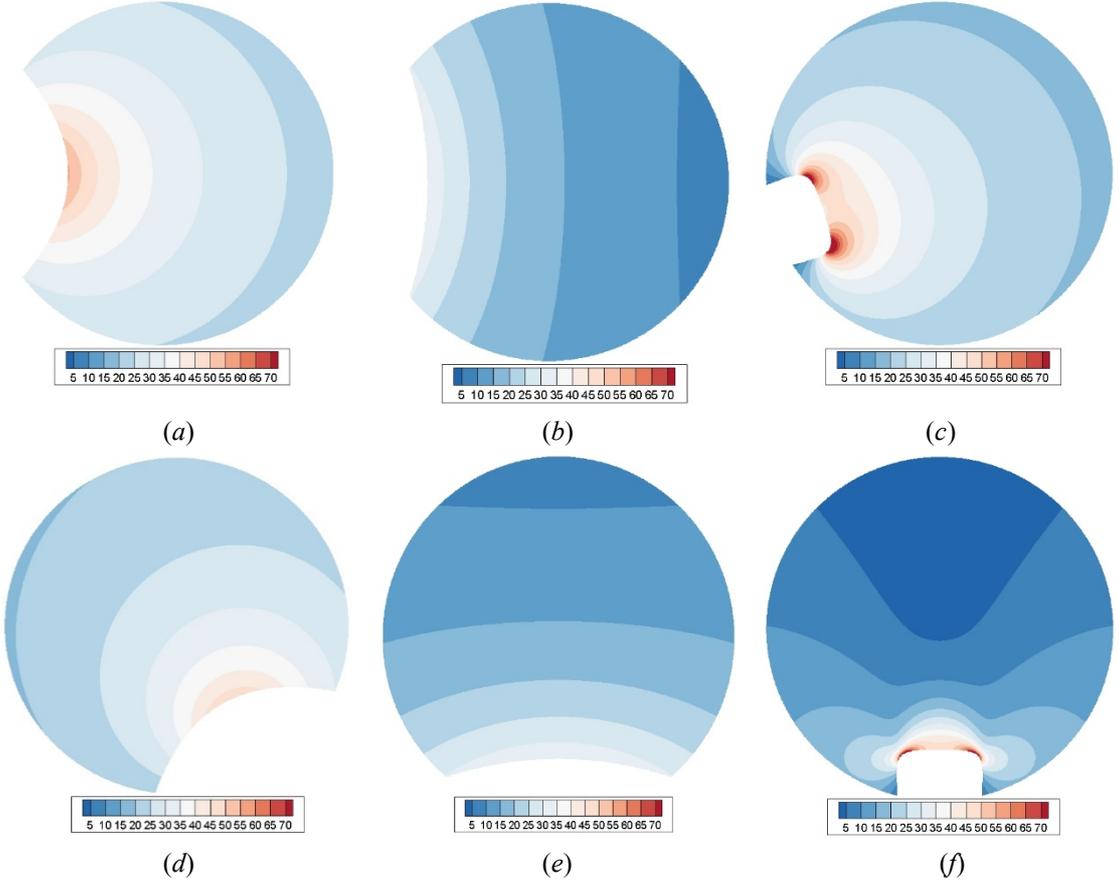

**Fig. 6.** Maximal shear stress scaled by $\sigma_0$ in (32) on the tips for three cavities: (*a-c*). Tip1 under uniform heat flux $\beta = 0$; (*d-f*). Tip2 under uniform heat flux $\beta = \pi/2$. The computational domain is a circular domain whose radius is one percent times $R$.

## 4.2 Change of toroidal normal stress (TNS) with the curvature of cavity contour

From Zou and He (2018), the curvature of the cavity contour can be calculated by

$$K_0 = \frac{1}{R|\phi'(\eta)|} \text{Re}\left[1 + \frac{\eta \phi''(\eta)}{\phi'(\eta)}\right]. \tag{50}$$

Considering that the boundary of the cavity is traction-free when the matrix is under the action of uniform heat flux at infinity, there is only toroidal normal stress (TNS) $\sigma_\vartheta$ on the boundary. Due to $\sigma_n = 0$ and $\sigma_\vartheta + \sigma_n =$

$\sigma_{11} + \sigma_{22}$, the relation (21)$_1$ yields the formula of toroidal normal stress as

$$\sigma_\vartheta = 4\text{Re}[\varphi'(t)] = 4\text{Re}\left[\frac{\varphi'(\eta)}{z'(\eta)}\right], \eta = e^{\iota\theta}. \tag{51}$$

It can be seen from Fig. 7 that in general the TNS varies sharply with the curvature of the cavity contour around the tips, but the TNSs around different tips have different variation characteristics and these characteristics might be affected by the direction of heat flux, even zero value of TNS appears at some tips. The tip with zero TNS depending on heat flux direction can be applied as the most ideal tip position for industrial design and material properties. It is worth mentioning that the curvature at the tips of pentagonal star presents bimodal characteristics, and the TNS is also different at the bimodal points (this phenomenon is caused by the platform at the tip). With the increase of terms in the Laurent polynomial of pentagonal star, the size of the platform at the tip will gradually decrease, and become infinitesimal when the pentagonal star is characterized by the Laurent series, and then the TNSs at the bimodal points will gradually approach to the same. For the case of bimodal tip, when the failure stress is reached, the crack may expand in both directions. According to Fig. 7, for three different shapes of cavities, it looks like that the maximum TNSs are obtained at the maximum curvature, and their values depends on the values of maximum curvature, so the maximum TNS are pentagonal star ($-99.073\sigma_0$)> triangle ($-55.75331\sigma_0$) > square ($-34.43897\sigma_0$).

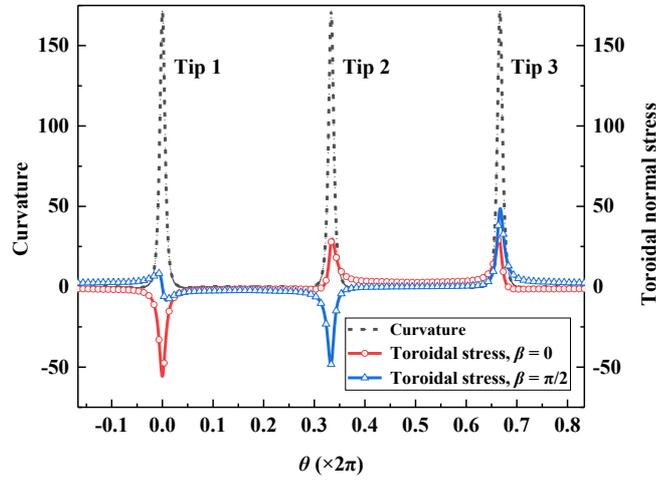

(a)

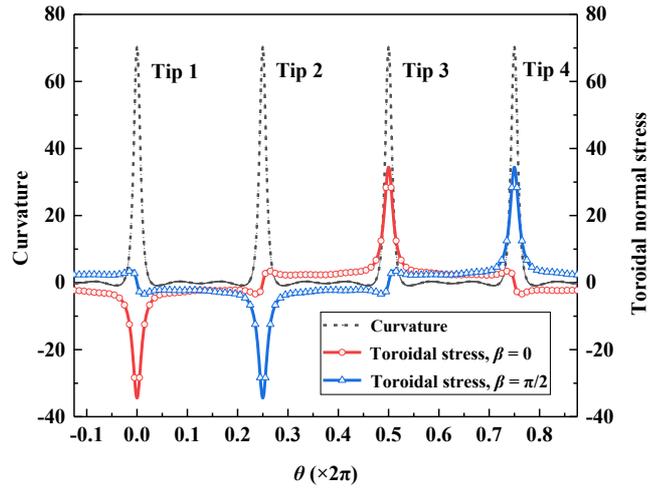

(b)

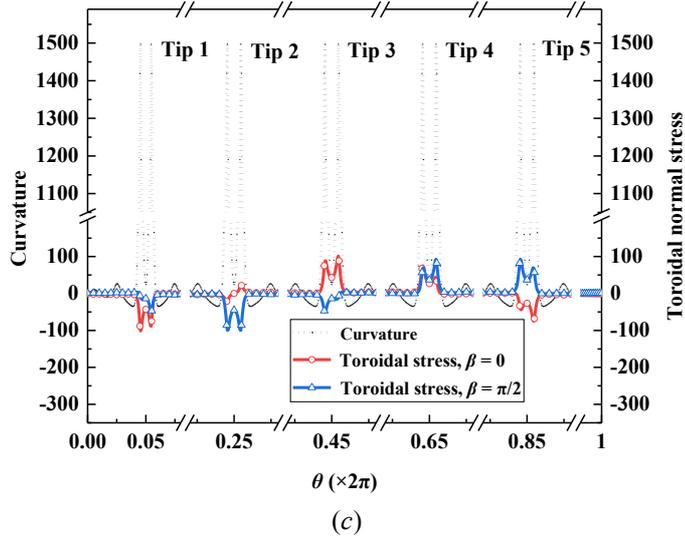

(c)

**Fig. 7.** Curvature scaled by $R^{-1}$ and toroidal normal stress scaled by $\sigma_0$ in (32) along the contour of the cavity: (a) triangle, (b) square, (c) pentagram.

### 4.3 Effect of heat flux direction on the maximal TNS around the tip

The direction of heat flux has a great influence on the stress distribution around the cavity. Here we are more concerned about the contour position where the stress reaches its extreme, and the heat flux direction when the stress reaches its extreme. Therefore, we only discuss the TNS at the maximum curvature point and the points with the maximal TNSs, with respect to the change of the heat flux direction. Since the stress concentration always happens around the tips, and the tips of regular polygons have the same configuration but different orientation, we choose Tip1 as a representative for discussion.

It is natural to use the external normal direction of the maximum curvature point of Tip1 as the reference direction. The slope of the normal at a point on the boundary has the formula

$$k_n = \frac{\text{Im}[\eta\phi'(\eta)]}{\text{Re}[\eta\phi'(\eta)]}, \tag{52}$$

depending on its coordinate $\eta$ in the image plane. Due to the existence of the tip platform in the case of pentagram, the external normal of the platform seems to be more suitable as the reference direction instead. The angle $\Theta$ between the direction of heat flux and the reference direction of Tip1 is taken from 0 to $2\pi$ anticlockwise in Fig. 8 and Fig. 9, while the TNS values are calculated from Eq. (51).

In Fig. 8, the TNS at the maximum curvature point and the maximal TNS on the contour are figured out for cavities of three shapes. In Fig. 9, the distance of the points, having the maximal TNSs, relative to the maximum curvature point is given. From these figures and the calculated data, we can list the following conclusions:

- For every point around the tip, there is a reference direction that the TNS at this point reaches its tension maximum when the heat flux direction is the same as this direction, and its compression maximum when the heat flux direction is opposite to this direction. The magnitude of the tension maximum is the same as that of the compression maximum. This reference direction is the external normal of the contour if this point is the symmetry point of the tip.
- For the cases of triangle and square, the maximum curvature point is also the symmetry point of the tip, and has the global maximal TNS when the heat flux direction is its external normal or opposite, as Fig. 8(a) shown. But when the heat flux direction is not parallel to its external normal, the point with the maximal TNS is no longer the maximum curvature point, as shown in Fig. 8(a), but deviates from the maximum curvature point with the distance becoming the largest when the heat flux direction is perpendicular to the normal, as shown in Fig. 9. For the special case of heat flux direction tangent to the contour, the TNS at the maximum curvature point becomes zero, and two points, with the maximum compression and the maximum tension, respectively, have maximal distance to the maximum curvature point, though the distance is smaller than two percent of *R*.

- For pentagram, the maximum curvature point is not the symmetry point of the tip, and doesn't have the global maximal TNS. The special feature is that the external normal of the platform is not a good reference direction, as shown in Fig. 8(a). From Fig. 8(b), we find that the direction of heat flux when the TNS reaches its global maximum is more suitable as a reference direction. And Fig. 9 shows the deviation of points with maximal TNS is smaller than 0.1% of *R* (the length of platform of the tip is $4.5677 \times 10^{-3}R$).

Therefore, as mentioned in Section 4.1, the point with the maximum stress value may not be the maximum curvature point, as the heat flux direction changes. Careful investigation shows that the maximal TNS doesn't occur at the maximum curvature point when the tip has no symmetric configuration. For pentagram characterized by (49.3), the contour point with the maximal TNS is about $1.007 \times 10^{-4}R$ from the maximum curvature point while the maximal TNS is -102.69848$\sigma_0$ (102.69848$\sigma_0$) in the heat flux direction of 10.95°(190.95°) degrees relative to the normal of platform of the tip, as comparison the maximal TNS at the maximum curvature point is -99.56408$\sigma_0$ (99.56408$\sigma_0$) in the heat flux direction of $-12.15°(167.85°)$ relative to the normal of platform of the tip.

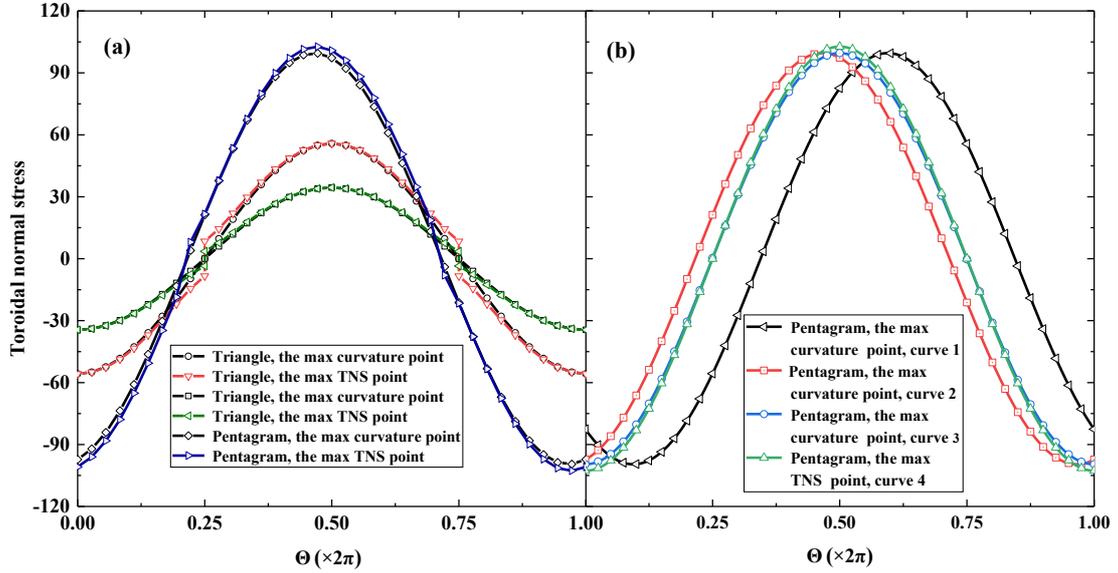

**Fig. 8** TNS of the maximum curvature point and the maximum TNS around Tip1 (all scaled by $\sigma_0$ of (32)), where Θ indicates the angle between the heat flux direction and the reference direction: (a). the reference direction is taken to be the normal of the maximum curvature point for triangle and square, but the normal of the platform for pentagram; (b). all for pentagram, curves 1-3 are TNS of the maximum curvature point using different reference directions, the external normal of the maximum curvature point for curve 1, the external normal of the platform for curve 2, and the direction of heat flux when the TNS reaches its global maximum for curve 3, curve 4 is the TNS of the point having the global maximum TNS with the same reference direction as curve 3.

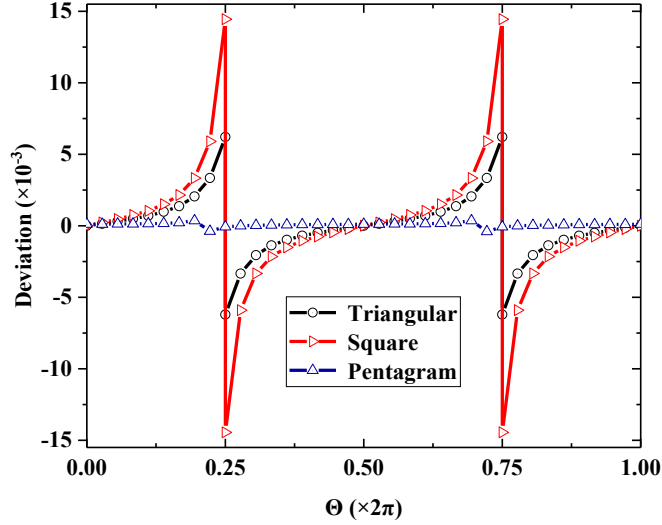

**Fig. 9** Deviation between the maximum stress point and the maximum curvature point around Tip1 with Θ being the angle between the heat flux direction and the reference direction.

### 4.4 Decay of hydrostatic pressure along lines starting from the tips

Along the direction of the symmetry axis of the cavity at the tip, take a straight-line segment with the length of 2 which is scaled by the radius $R_0$ of circumscribed circle of the cavity shape, starting from the maximum curvature point of the tip, the hydrostatic pressure (47) on the line segments near Tip1s of the three cavities are shown in Fig. 10(*a*)-(*c*). It can be seen that except that the hydrostatic stress on the line is always 0 when the tip has normal perpendicular to the heat flux direction (pentagram has an instantaneous change due to the nonsymmetric tip), the hydrostatic pressure on the line decreases rapidly and gradually in the distance, and the attenuation rate gradually slows down with the increase of distance. The effect of heat flow direction on hydrostatic pressure does not change with the distance, and the stress decay rate at the tip of a cavity with different shapes is also positive correlation with the curvature: pentagram (from $99.81\sigma_0$ to $1.69\sigma_0$) > triangle (from $55.78\sigma_0$ to $2.11\sigma_0$) > square (from $34.44\sigma_0$ to $2.21\sigma_0$), when the segments are parallel to the heat flux direction.

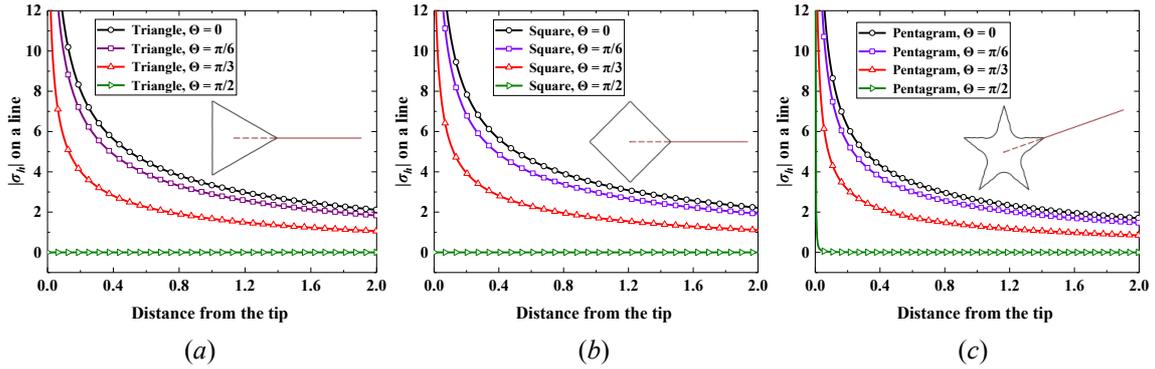

**Fig. 10.** Hydrostatic pressure $|\sigma_h|$ scaled by $\sigma_0$ in (32) on a line at the tips: (*a*) triangle, (*b*) square, (*c*) pentagram.

## 5 Conclusion

In this paper, the two-dimensional thermoelastic problem of infinite medium with a cavity subjected to uniform heat flux is studied by using the plane complex variable theory. In the aspect of solving the problem, we obtain the explicit analytic solutions of K-M potentials and relative rigid-body translation by equivalence analysis tactics in series expansion. By comparison, we find that the previous solutions containing positive power terms have the operation problem of big numbers when the field points are away from the boundary of the cavity, especially for the cavity with complicated shape requiring coefficients of high degrees in the description of Laurent polynomials. the in the existing literature.

The maximum stress plays a key role in engineering design and material properties. Using the new solution for a cavity with arbitrary shape, with triangle, square and pentagonal stars as representative shapes, we study the thermal stress distribution by investigating the effects of different cavity shapes, at different tips and under different heat flux directions. The major conclusions can be listed as follows:

- The thermal stress exists around the cavity and the stress concentration is obvious at the tips. For regular polygon, the stress distribution behaves some symmetry according to the symmetry of the cavity shape. The maximum stress is positively correlated with the maximal curvature. For different cavity shapes scaled by the area, the stress concentration becomes serious as: pentagonal star > triangle > square.
- When the heat flux direction is parallel to the external normal direction of the tip, the stress at this point reaches the maximum; when the heat flow direction is perpendicular to the normal direction of the tip, the stress at this point reaches the minimum; When there is an angle between the heat flow direction and the external normal direction of the tip, the maximum stress point and the maximum curvature point will also have a small deviation;
- With the increase of the distance from the tip, the hydrostatic pressure decreases rapidly, and its rate is also proportional to the curvature.

## Disclosure statement

No potential conflict of interest was reported by the authors.

## Funding

This work is supported by the National Science Foundation of China (NSFC), grant no. 11962017.

## Appendix A. Detailed derivation of explicit analytical solution

According to the analysis of Zou and He (2018), all terms in (35) can be expanded in terms of the power of $\eta$, and the Cauchy integral formula will guarantee that the part of non-positive powers must be in balance everywhere. Thus, the terms of positive powers can be omitted and an equivalent relation operator "$\sim$" can be introduced to carry out the potentials.

First of all, by defining a function of non-positive powers

$$G(\eta^{-1}) \equiv \sum_{k=0}^{N} g_k \eta^{-k} \sim \frac{\phi(\eta)}{\phi'(\eta)} = \left(1 - \sum_{l=1}^{N} l\bar{b}_l \eta^{l+1}\right)^{-1} \sum_{k=1}^{N} b_k \eta^{-k}, \tag{A.1}$$

we have the equivalent relation

$$\sum_{k=0}^{N} g_k \eta^{-k} \sim \sum_{k=0}^{N-2}\left(\sum_{l=1}^{N-k-1} lg_{k+l+1}\bar{b}_l\right)\eta^{-k} + \sum_{k=1}^{N} b_k \eta^{-k} \tag{A.2}$$

to achieve

$$g_N = b_N, \qquad g_{N-1} = b_{N-1}, \tag{A.3.1}$$

and the recursive formulae between $g_k$ and $b_k$

$$g_{N-k} = b_{N-k} + \sum_{l=1}^{k-1} lg_{N-k+l+1}\bar{b}_l, \quad k = 2, \ldots, N-1; \; g_0 = \sum_{l=1}^{N-1} lg_{l+1}\bar{b}_l. \tag{A.3.2}$$

Based on this, we have

$$\frac{\phi(\eta)}{\phi'(\eta)}\overline{\varphi'_p(\eta)} \sim -R \sum_{n=2}^{N} g_k \eta^{-k} \sum_{l=1}^{\infty} l\bar{\alpha}_l \eta^{l+1} \sim -R \sum_{n=0}^{N-2} \eta^{-n} \sum_{l=1}^{N-n-1} l\bar{\alpha}_l g_{n+l+1}. \tag{A.4}$$

and the non-positive part of the right of (35)

$$R \sum_{k=0}^{N-1} c_k \eta^{-k} = -2\mu U_0 - \frac{\bar{A}\phi(\eta)}{\phi'(\eta)}\eta \sim -2\mu U_0 - \bar{A}\sum_{k=1}^{N} g_k \eta^{-k+1}. \tag{A.5}$$

Namely

$$c_0 = -\bar{\sigma}g_1 - \frac{2\mu}{R}U_0; \; c_k = -\bar{\sigma}g_{k+1}, k = 1, 2, \ldots, N-1. \tag{A.6}$$

Then, the equivalent form of (35)

$$\varphi_p(\eta) + \frac{\phi(\eta)}{\phi'(\eta)}\overline{\varphi'_p(\eta)} + \overline{\psi_p(\eta)} \sim -2\mu U_0 - \frac{\bar{A}\phi(\eta)}{\phi'(\eta)}\eta \tag{A.7}$$

shows that the potential $\varphi_p(\eta)$ must have maximal negative power $N-1$, and the equation

$$\sum_{n=1}^{N-1} \alpha_n \eta^{-n} - \sum_{n=0}^{N-2} \eta^{-n} \sum_{l=1}^{N-n-1} l\bar{\alpha}_l g_{n+l+1} = \sum_{n=0}^{N-1} c_n \eta^{-n} \tag{A.8}$$

yields the linear equations

$$\alpha_k = c_k + \sum_{l=1}^{N-k-1} l\bar{\alpha}_l g_{k+l+1}, \qquad k = 1, 2, \ldots, N-1 \tag{A.9}$$

for the undetermined coefficients $\{\alpha_k, k = 1, \cdots, N-1\}$ and the formula of rigid-body translation $U_0$

$$2\mu U_0 = R\sum_{l=1}^{N-1} lg_{l+1}\bar{\alpha}_l - \bar{A}g_1. \tag{A.10}$$

Zou and He (2018) also proposed a recursive solution method for the linear equations (A.9). Starting with

$$\alpha_k^{(0)} = c_k, k = 1, 2, \ldots, N-1, \tag{A.11}$$

and doing

$$\alpha_k^{(n+1)} = c_k + \sum_{l=1}^{N-k-1} lg_{k+l+1}\bar{\alpha}_l^{(n)}, \qquad n = 0, 1, 2, \ldots \tag{A.12}$$

to achieve the result of convergence.

For another potential $\psi_p(\eta)$, conjugating the equation (35) and then multiplying it by $\phi'(\eta)$ gives

$$\phi'(\eta)\overline{\varphi_p(\eta)} + \overline{\phi(\eta)}\varphi'_p(\eta) + \phi'(\eta)\psi_p(\eta) = -2\mu\overline{U}_0\phi'(\eta) - A\overline{\phi(\eta)}\eta^{-1}. \tag{A.13}$$

From the equivalent relations

$$\phi'(\eta)\overline{\varphi_p(\eta)} \sim -R\sum_{k=1}^{N-1}kb_k\bar{\alpha}_{k+1} - R\sum_{l=1}^{N}\eta^{-l}\sum_{k=l}^{N}kb_k\bar{\alpha}_{k-l+1},$$

$$\overline{\phi(\eta)}\varphi'_p(\eta) \sim \eta^{-1}\varphi'_p(\eta) - R\sum_{k=1}^{N-1}k\alpha_k\bar{b}_{k+1} - R\sum_{l=1}^{N}\eta^{-l}\sum_{k=l}^{N}k\alpha_k\bar{b}_{k-l+1},$$

and the right terms of (A.13), it can be judged that the maximal negative power of $\phi'(\eta)\psi_p$ is $N+1$, and yield

$$\phi'(\eta)\psi_p = R\sum_{k=1}^{N}\lambda_k\eta^{-k} - \eta^{-1}\varphi'_p(\eta) - A\eta^{-2} + 2\mu\overline{U}_0[1 - \phi'(\eta)], \tag{A.14}$$

with

$$\lambda_k = \sum_{l=k}^{N}l(b_l\bar{\alpha}_{l-k+1} + \alpha_l\bar{b}_{l-k+1}), \quad k = 1, \dots, N. \tag{A.15}$$

The residual of the constant term viewed from the right is

$$C = R\sum_{k=1}^{N-1}k(b_k\bar{\alpha}_{k+1} + \alpha_k\bar{b}_{k+1}) - 2\mu\overline{U}_0 - A\bar{b}_1. \tag{A.16}$$

**Note**: Using (A.10) and $g_1 = b_1 + \sum_{l=1}^{N-2}lg_{l+2}\bar{b}_l$ from (A.3.2), we have

$$-2\mu\overline{U}_0 - A\bar{b}_1 = A\bar{g}_1 - R\sum_{l=1}^{N-1}l\bar{g}_{l+1}\alpha_l - A\bar{b}_1 = A\sum_{l=1}^{N-2}l\bar{g}_{l+2}b_l - R\sum_{l=1}^{N-1}l\bar{g}_{l+1}\alpha_l. \tag{A.17}$$

Substituting it into (A.16) yields

$$\begin{aligned}
C &= R\sum_{k=1}^{N-1}k(b_k\bar{\alpha}_{k+1} + \alpha_k\bar{b}_{k+1}) + A\sum_{k=1}^{N-2}k\bar{g}_{k+2}b_k - R\sum_{k=1}^{N-1}k\bar{g}_{k+1}\alpha_k \\
&= R\sum_{k=1}^{N-1}k[b_k(\bar{\alpha}_{k+1} + \sigma\bar{g}_{k+2}) + \alpha_k(\bar{b}_{k+1} - \bar{g}_{k+1})] \\
&= R\sum_{k=1}^{N-1}k\left(b_k\sum_{l=1}^{N-k}l\alpha_l\bar{g}_{k+l} - \alpha_k\sum_{l=1}^{N-k}l\bar{g}_{k+l}b_l\right) \equiv 0,
\end{aligned} \tag{A.18}$$

where uses are made of the relations (A.3.2), (A.6)₁ and (A.9).